\documentclass[12pt]{article}
\usepackage{graphicx}
\begin{document}
Computer Simulations for Biological Ageing and Sexual Reproduction
\bigskip

D. Stauffer$^1$, P.M.C. de Oliveira, S. Moss de Oliveira, T.J.P. Penna and
J.S. S\'a Martins$^2$
\bigskip

Instituto de F\'{\i}sica, Universidade Federal Fluminense, 
  
Av. Litor\^anea s/n, Boa Viagem, Niter\'oi 24210-340, RJ, Brazil:

\medskip
$^1$visiting from 
Inst. for Theoretical Physics, Cologne University, D-50923 K\"oln, Euroland

\medskip
$^2$  Colorado Center for Chaos and Complexity/CIRES, 

University of Colorado, Boulder CO 80309, USA

\bigskip

	The sexual version of the Penna model of biological ageing,
simulated since 1996, is compared here with alternative forms of
reproduction as well as with models not involving ageing. In particular we
want to check how sexual forms of life could have evolved and won over
earlier asexual forms hundreds of million years ago. This computer model
is based on the mutation-accumulation theory of ageing, using bits-strings
to represent the genome. Its population dynamics is studied by Monte Carlo
methods.

\medskip
Keywords: parthenogenesis, genome, menopause, testosteron, 
Monte Carlo simulation

\newpage
\section{Introduction}

        Can physicists contribute to understand biological subjects? Since
the first attempts by the Nobel laureate Schr\"odinger (1944), there were
a lot of tentative answers to this question, probably most of them
useless. What particular knowledge can physicists bring to Biology? One
particular tentative, biased answer for this second question is presented
below. It is biased because it concerns just the authors' traditional line
of research.

        Critical phenomena appear in macroscopic physical systems
undergoing continuous phase transitions. An example is water crossing the
critical temperature of $374^{\rm o}\,$C, above which one can no longer
distinguish liquid from vapour. Another is a ferromagnetic material which
loses its spontaneous magnetisation when heated above its critical
temperature. Such systems present unusual behaviours. For instance, some
quantities increase without limits as one approaches more and more the
critical point, as the water compressibility: a very small pressure leads
to an enormous volume shrinkage. Analogously, by applying a very small
magnetic field, one can drastically increase the magnetisation of a
ferromagnetic sample. In both examples, also the specific heat diverges at
the critical point, meaning that the system can absorb or deliver a large
amount of heat, without any sensible temperature variation. Needless to
mention the important practical applications of such a behaviour,
according to which a fine tuning of some quantity can lead to an enormous
variation of another related quantity. All modern electronics, for
instance, is based on the possibility of getting an electric current
passing through an otherwise insulating device, simply by applying a small
electric field. Also, some plastic materials can undergo very large volume
expansions under very small electric or magnetic impulses: they are used
for manufacturing of artificial muscles, catheters which unblock arteries,
microengines, etc.

        These features have attracted the attention of physicists since
more than a century. They discovered an also unusual behaviour concerning
the mathematical description of such systems: the appearence of {\bf
power-laws}, i.e. $Q \sim \vert T-T_{\rm c} \vert^{-\gamma}$ or $C \sim
\vert T-T_{\rm c} \vert^{-\alpha}$, where $Q$ is the quoted diverging
quantity (compressibility or magnetic susceptibility), $C$ is the specific
heat, and $\vert T-T_{\rm c} \vert$ measures how far the system is from
its own critical point. The symbol $\sim$ represents proportionality.
Critical exponents like $\gamma$, $\alpha$, etc are characteristic of the
corresponding quantity, $Q$, $C$, etc.

        The most interesting feature of these phenomena is the so-called
universality: the precise values of the exponents $\gamma$, $\alpha$, etc
are the same for entire classes of completely different systems. For
instance, $\alpha = 0.12$ for both water and any ferromagnet in which
magnetisation presents uni-axial symmetry. Also, $\gamma = 1.24$ for both
the water compressibility and the magnetic susceptibility of the
ferromagnetic material. Besides critical exponents, many other qualitative
and quantitative characteristics of the various systems belonging to the
same universality class coincide as well. In spite of having been observed
much before, these coincidences remained unexplained until the work of
Wilson (1971), three decades ago, who was awarded with the Nobel prize
because of this work (see also: Wilson \& Kogut 1974; Wilson 1979). The
key concept needed to understand this phenomenon is the decaying of
correlations with increasing distances. Suppose one picks two points
inside the system, separated by a distance $x$. How much a perturbation
performed at one of these points will be felt at the other? The
correlation $I$ between these two points is a measure of this mutual
influence, and generally decays for larger and larger values of $x$,
according to the exponential behaviour

$$I \sim \exp{(-x/\xi)}\,\,\,\,\,\,\,\,\,\,\,\,\,\,\,\,\,\,\,\,\,\,\,\,
{\rm (for\,\, non\,\, critical\,\, situations)},\eqno({\rm NC})$$

\noindent where $\xi$ is the so-called correlation length. Would one take
two points distant from each other a distance $x$ larger than $\xi$, the
correlation $I$ would be negligible. This means that one does not need to
study the macroscopic system as a whole, with its enormous number of
component units: it is enough to take a small piece of the system with
linear dimensions of the same order as $\xi$ (for instance, a sphere with
radius, say, $10\xi$). Once one knows, for instance, the specific heat of
this small piece, that of the whole system is obtained by a simple volume
proportionality $C \sim V$ or $Q \sim V$.

        However, the nearer the system is to its critical point, the
larger is $\xi$, and the larger is the ``small'' piece representing the
whole, i.e. $\xi \sim \vert T-T_{\rm c} \vert^{-\nu}$. Just {\bf at} the
critical point, one can no longer break the system into small pieces: the
macroscopic critical behaviour of the system is no longer proportional to
its volume. Instead, critical quantities become non-linear, non-extensive,
and behave as $Q \sim V^{\Phi_\gamma}$ or $C \sim V^{\Phi_\alpha}$, where
$\Phi_\gamma = \gamma/3\nu$, $\Phi_\alpha = \alpha/3\nu$, etc. Also, the
above exponential form (NC) for $I$ concerns only the dominating decay
valid for a finite $\xi$. {\bf At} the critical point where $\xi \to
\infty$, however, other sub-dominating terms enter into the scene, i.e.

$$I \sim x^{-\eta}\,\,\,\,\,\,\,\,\,\,\,\,\,\,\,\,\,\,\,\,\,\,\,\,
{\rm (for\,\, critical\,\, situations)},\eqno({\rm C})$$

\noindent where $\eta$ is another critical exponent.

        Both forms (NC) and (C) mean that correlations decay for larger
and larger distances. The important conceptual difference is that in (NC)
they decay much faster, according to a characteristic length scale $\xi$
above which correlations become negligible. On the opposite, there is no
characteristic length scale in the critical case (C): correlations are
never negligible even between two points very far from each other, inside
the system. Thus clusters and holes are observed at all sizes, a crucial
property e.g. for electrophoresis.

	This is a big mess for theoretical physicists: since any tentative
to break such systems into small pieces is denied, they are forced to
treat them as a whole. Fortunately, this same strange non-linear, critical
behaviour leads to another very important property: most microscopic
details of the system are irrelevant in what concerns its critical
behaviour, since large distances dominate the scenario. That is why water
compressibility presents exactly the same critical exponent $\gamma$ as
the magnetic susceptibility of any uni-axial ferromagnet, as well as the
same values for the other exponents $\alpha$, $\nu$, $\eta$, etc, and thus
the same critical behaviour. Not only water and such ferromagnetic
materials, but also any other natural or artificial system which belongs
to the same (huge) universality class. One example of such mathematical
toys is the famous Ising model: each point on a regular lattice holds a
binary variable (a number 0 or 1), and interacts only with its neigbouring
sites. No movement at all, no molecules, no atoms, no electrons
interacting through complicated quantum rules. The only similarities
between this toy model and real water are two very general ingredients:
the three-dimensionality of the space and the one-dimensionality of the
main variables involved (the numbers 0 or 1 within the toy, and the
liquid-vapour density difference within water, also a number, as opposed
to a three-dimensional vector).  Nevertheless, one can use this very
simple toy in order to obtain the critical behaviour common to all much
more complicated systems belonging to the same universality class.

        However, even the study of these toy models is far from trivial,
due to the already quoted impossibility of breaking the system into small,
separate pieces. Thus, the main instrument is the computer, where one can
store the current state of each unit, i.e. a number 0 or 1, into a single
bit of the memory. By programming the computer to follow the evolution of
this artificial system time after time, i.e. by repeatedly flipping 0s
into 1s (or vice-versa) according to some prescribed microscopic rule, one
can measure the various quantities of interest. Note that this approach
has nothing to do with the numerical solution of a well posed mathematical
problem defined by specific equations. Instead, the idea is {\bf to
simulate} the real {\bf dynamical} behaviour of the system on the
computer, and {\bf to measure} the interesting quantities. During the last
half century, this ``almost experimental'' technique was tremendously
developed by the (now-called) computational physicists, a fast growing
scientific community to which the authors belong (Stauffer \& Aharony
1994; de Oliveira 1991; Moss de Oliveira et al. 1999).

        Biological evolution (Darwin 1859) also presents the same
fundamental mathematical ingredient which characterise physical critical
systems: the power-laws. A lot of evidences are known, today (see, for
instance, Kauffman 1993 and 1995; Bak 1997). A simple and well known
example is the number $A$ of still alive lineages within an evolving
population: it decays in time according to the power-law $A \sim t^{-1}$,
where the exponent $-1$ can be exactly obtained from the coalescence
theory (see, for instance, Excoffier 1997). According to this, after many
generations, all individuals of the population are descendents of a single
lineage-founder ancestor. The number of generations one needs to wait for
this coalescence is proportional to the number of founder individuals, due
to the value $-1$ of the exponent. Also, during the whole evolution of the
population, the number $E$ of already-extinct lineages with $n$
individuals behaves as $E \sim n^{-0.5}$, where the new exponent $-0.5$ is
also exactly known. The interesting point is that these exponents are {\bf
universal}, i.e. their precise values do not change for different
microscopic rules dictating how individuals die, how they are born, etc.
Another simple example is the evolution of a recessive disease: the
frequency of the recessive gene among the evolving population also decays
in time as a power-law, thus {\bf without a characteristic extinction
time}. Due to this particular mathematical decaying feature, the recessive
gene extinction is postponed forever (Jacquard 1978). An explanation for
the narrow relation between biological evolution and critical dynamics is
presented by de Oliveira (2000).

        The Penna model for biological ageing (Penna 1995) is entirely
based on Darwinian evolution, and may be compared with the Ising model,
for this particular evolutionary phenomenon: genes are also represented by
binary variables (0 for ordinary genes, 1 for harmful ones). It spreads
widely during the last half decade, and was applied to many different
biological problems involving ageing, always within the general
interpretation above: {\bf a very simple model supposed to reproduce the
universal features of much more complicated, real phenomena.}

	Senescence, or biological ageing, can mean many things; for
computer simulations it is best defined as the increase of mortality with
increasing age. It seems not to exist for bacteria, where even the concept
of death is difficult to define, but for humans as well as for other
organisms (Vaupel et al. 1998)  this rapid increase of the probability to
die, after childhood diseases are overcome, is well known. Fig.1 shows
typical human data for a rich country.

	The reasons for ageing are controversial (Watcher \& Finch 1997;
see also the whole special issues of {\it La Recherche}: July/August 1999
and {\it Nature}: November 9th, 2000). There may be exactly one gene for
longevity, or senescence comes from wear and tear like for insect wings
and athlete's limbs, from programmed cell death (apoptosis - Holbrook et
al. 1996), from metabolic oxygen radicals destroying the DNA (see for
instance Azbel 1994), or from mutation accumulation (Rose 1991). The
computer simulations reviewed here use this last assumption, which does
not exclude all the other reasons. For example, the oxygen radicals may
produce the mutations which then accumulate in the genome transmitted from
one generation to the next. Except if stated otherwise, the mutations here
are all detrimental and inherited.

	After a short description of the model in section 2, we deal in
section 3 with the question whether sexual reproduction was better or
worse than asexual reproduction hundreds of million years ago when sex
appeared, while section 4 tries to explain why today's women live longer
than men and have menopause.  Section 5 reviews other aspects, and Section
6 gives a short summary.

	A more detailed account, but without the results of 1999 and 2000
emphasized here, is given in our book (Moss de Oliveira et al. 1999).

\section{The Penna model}

	In the original asexual version of the Penna model (Penna 1995)  
the genome of each individual is represented by a computer word
(bit-string) of 32 bits (each bit can be zero or one). It is assumed that
each bit corresponds to one ``year'' in the individual lifetime, and
consequently each individual can live at most for 32 ``years''. {\it A bit
set to one means that the individual will suffer from the effects of a
deleterious inherited mutation (genetic disease) in that and all following
years}. As an example, an individual with a genome $10100...$ would start
to become sick during its first year of life and would become worse during
its third year when a new disease appears.  In this way the bit-string
represents in fact a ``chronological genome''.  The biological motivation
for such a representation is, for instance, the Alzheimer disease: its
effects generally appear at old ages, although the corresponding defective
gene is present in the genetic code since birth.

	The extremely short size of the 32 bit-string used in the model
would be totally unrealistic if all our genes were related to
life-threatening diseases. However, among the average number of $10^8$
units we have in our real genome, only around $10^4$ to $10^5$ units play
a functional role. Moreover, only a subgroup of these will give rise to a
serious disease at some moment of the individual lifetime. Besides,
qualitatively there was no difference when 32, 64 and 128 bits were taken
into account (Penna \& Stauffer 1996).

	One step of the simulation corresponds to reading one bit of all
genomes. Whenever a new bit of a given genome is read, we increase by one
the individual's age.  The rules for the individual to stay alive are: 1)
The number of inherited diseases (bits set to 1) already accumulated until
its current age must be lower than a threshold $T$, the same for the whole
population. In the example given above, if $T=2$ the individual would live
only for 2 years. 2) There is a competition for space and food given by
the logistic Verhulst factor $V=1-N(t)/N_{max}$, where $N_{max}$ is the
maximum population size the environment can support and $N(t)$ is the
current population size. We usually consider $N_{max}$ ten times larger
than the initial population $N(0)$. At each time step and for each
individual a random number between zero and one is generated and compared
with $V$: if it is greater than $V$, the individual dies independently of
its age or genome. The smaller the population size is, the greater is the
probability of any individual to escape from this random killing factor.

	If the individual succeeds in staying alive until a minimum
reproduction age $R$, it generates $b$ offspring in that and all following
years (unless we decide to set also some maximum reproduction age). The
offspring genome is a copy of the parent's one, except for $M$ randomly
chosen mutations introduced at birth. Although the model allows good and
bad mutations, generally we consider only the bad ones. In this case, if a
bit 1 is randomly tossed in the parent's genome, it remains 1 in the
offspring genome; however, if a bit zero is randomly tossed, it is set to
1 in the mutated offspring genome. In this way, for the asexual
reproduction the offspring is always as good as or worse than the parent.
Even so, a stable population is obtained, provided the birth rate $b$ is
greater than a minimum value, which was analytically obtained by Penna \&
Moss de Oliveira (1995). In fact, the population is sustained by those
cases where no mutation occurs, when a bit already set to 1 in the parent
genome is chosen.  These cases are enough to avoid mutational meltdown,
that is, extinction due to accumulation of deleterious mutation, first
considered by Lynch \& Gabriel (1990). The reason why we consider only
harmful mutations is that they are 100 times more frequent than the
backward ones (reverse mutations deleting harmful ones - Pamilo et al.
1987).

	The sexual version of the Penna model was first introduced by
Bernardes (1995 and 1996), followed by Stauffer et al. (1996)  who adopted
a slightly different strategy. We are going to describe and use the second
one (see also Moss de Oliveira et al. 1996).  Now individuals are
diploids, with their genomes represented by two bit-strings that are read
in parallel.  One of the bit-strings contains the genetic information
inherited from the mother, and the other from the father. In order to
count the accumulated number of mutations and compare it with the
threshold $T$, it is necessary to distinguish between recessive and
dominant mutations. A mutation is counted if two bits set to 1 appear at
the same position in both bit-strings (inherited from both parents) or if
it appears in only one of the bit-strings but at a dominant position
(locus).  The dominant positions are randomly chosen at the beginning of
the simulation and are the same for all individuals.

	The population is now divided into males and females. After
reaching the minimum reproduction age $R$, a female randomly chooses a
male with age also equal to or greater than $R$ to breed (for sexual
fidelity see Sousa \& Moss de Oliveira 1999).  To construct one offspring
genome first the two bit-strings of the mother are cut in a random
position (crossing), producing four bit-string pieces. Two complementary
pieces are chosen to form the female gamete (recombination). Finally,
$m_f$ deleterious mutations are randomly introduced. The same process
occurs with the male's genome, producing the male gamete with $m_m$
deleterious mutations. These two resulting bit-strings form the offspring
genome.  The sex of the baby is randomly chosen, with a probability of
$50\%$ for each one.  This whole strategy is repeated $b$ times to produce
the $b$ offspring. The Verhulst killing factor already mentioned works in
the same way as in the asexual reproduction.

	A very important parameter of the Penna model is the minimum
reproduction age $R$. According to mutation accumulation-theory, Darwinian
selection pressure tries to keep our genomes as clean as possible until
reproduction starts. For this reason we age:  mutations that appear early
in life are not transmitted and disappear from the population, while those
that become active late in life when we barely reproduce can accumulate,
decreasing our survival probability but without risking the perpetuation
of the species.  One of the most striking examples of such a mechanism is
the catastrophic senescence of the pacific salmon and other species called
semelparous: In these species all individuals reproduce only once in life,
all at the same age. This can be easily implemented smply by setting a
maximum reproduction age equal to $R$. After many generations, the
inherited mutations have accumulated in such a way that as soon as
reproduction occurs, individuals die. This explanation was given by Penna
et al. (1995), using the Penna model (see also Penna \& Moss de Oliveira
1995 and a remark from Tuljapurkar on page 70 in Wachter \& Finch 1997).
\bigskip

\section{Comparison of sexual and asexual reproduction}
\medskip

\subsection{Definitions}
\medskip

	In this section we check which way of reproduction is best:
Sexual, asexual or something in between. We denote as asexual (AS) and
sexual (SX) the simulation methods described in the previous section, that
means cloning of a haploid genome for AS, and crossover for diploid
genomes with males and females separated for SX. Intermediate
possibilities which will also be compared are apomictic parthenogenesis
(AP), meiotic parthenogenesis (MP), hermaphroditism (HA), and mixtures of
them. One could also group AS, AP and MP into asexual and HA and SX into
sexual reproduction. Parasex, the exchange of haploid genome parts between
different bacteria, is not simulated here.

	To find out which way is the most successful one we simulate each
choice separately with the same parameters, in particular with the same
$N_{max}$ for the Verhulst factor taking into account the limits of space
and food. The choice with the largest equilibrium population, after the
initial transient phenomena are overcome, is regarded as the best. We
assume it would win in a Darwinian selection (see Stauffer et al. 2000 for
some justification) if different populations following these different
ways of reproduction would compete against each other in the same
environment, without any symbiosis or predator-prey relation between them.

	AS and SX were defined already in the preceding section. For AP
the diploid genome is copied without crossover, only mutations. For MP the
diploid genome is crossed over, and one of the two resulting haploid
bit-strings is randomly chosen, duplicated and mutated to form the new
diploid genome. HA is similar to SX except that there is no separation
into males and females; instead all of them can generate offspring and
each individual selects randomly a partner from the whole population to
exchange genome as in SX. Fig.2 summarizes the four versions
schematically.

	In all the copying of genome (bit-strings), point mutations are
assumed to happen with the same probability per bit. Thus typically for AS
with a genome of 32 bits we assume one mutation per generation for the
whole genome, while for the diploid cases AP, MP, HA and SX we assume two.
(We assume the same mutation rate for males and females in the Penna model
simulations.) The birth rate is also assumed to be the same for all
birth-giving individuals. For example, for AS, AP, MP and HA we have four
offspring per suitable individual and per year, while for SX we have four
offspring per suitable female and per year. Thus the birth rate averaged
over males and females is only two instead of four. And we have to find
out whether this loss of a factor of two in the average birthrate for SX
is overcome by advantages not contained in the other ways of reproduction.

	(For HA as simulated by Stauffer et al. (2000) during one
iteration some individuals have already aged by one ``year'', while others
have not yet aged. Results do not change much, see topmost data in Fig.3
below, if now in one iteration we first let everybody age one time unit,
and only afterwards partners are selected.)

\subsection{Comparison without ageing}

	The Redfield model (Redfield 1994) is an elegant model requiring
much less computer time than the Penna model, but having no age structure.
It is not a population dynamics model following the lifetime of each
individual, but only simulates their probabilities to survive up to
reproduction. The mortality increases exponentially with the number of
mutations in the individual. For the sexual variant the number of
mutations in the child is determined by a binomial distribution such that
on average the child has as its own number of mutations half the number of
the father, plus half the number from the mother. At birth, new mutations
are added following a Poisson distribution, for both AS and SX. Because of
the lack of an explicit genome, the forms AP, MP, HA between AS and SX
were not simulated.

	This model triggered many publications in the physics literature
since it originally made SX much worse than AS: The average mortality was
about 25 percent for AS, and did not change when the simulation switched
to SX. But still the males were eating the food away from the females.  
Actually, the male mutation rate is much higher than the female one, and
when that was taken into account the mortality with SX became much higher
than for AS (Redfield 1994).

	However, the picture changed drastically when we took into account
(Stauffer et al. 1996) that most hereditary diseases are recessive (acting
only when both father and mother had them in their transmitted genome) and
not dominant (acting already when only one of the two inherited
bit-strings has them). Then the mortality decreased by about an order of
magnitude, and SX became much better than AS. The same drastic improvement
was found when for SX the females selected only males with few mutations.

	However, since the males do not give birth, these simulations
(Stauffer et al. 1996)  required the female birth rate for SX to be twice
as high as the AS birth rate. Correcting this factor of two, and still
assuming that only 20\% of the mutations are dominant diseases, SX lost
out for the originally selected mutation rate (Redfield 1994) of 0.3.
Increasing the mutation rate to 1 for both AS and SX, SX won (Stauffer
1999) over AS if the male mutation rate was the same as the female one,
and AS won over SX when the male mutation rate was three or more times
higher than the female one. Thus, as also observed in Nature, sometimes
asexual and sometimes sexual reproduction is better.

	A more realistic model, involving an explicit genome in the form
of bit-strings, was more recently investigated by \"Or\c cal et al.
(2000). It did not involve ageing, however, since all bit positions were
treated equally. Instead, \"Or\c cal et al. (2000) used the Jan et al.
(2000) parameter $\mu$ defined such that only individuals with $\mu$ and
more mutations exchange genome. (The model is then closer to HA than to
SX.) Healthy individuals without many mutations reproduce similarly to AS.
Five different versions were studied, depending on the number of offspring
and on whether individuals with $\mu$ and more mutations mate only with
each other or also with a wider population having less mutations. The
simulation showed that in none of the five cases the sexual population
died out; in one case it won even completely and made the AS population
extinct.

	These two models (Redfield 1994 and \"Or\c cal et al. 2000)  thus
give the same result: The simpler asexual way of life is not necessarily
better than the more complicated way of genome exchange. And this possible
justification of sex comes only from intrinsic genetic reasons, not from
extrinsic or social reasons like parasites, changing environment, or child
protection. On the other hand, there are also cases were asexual
reproduction is better. With ageing included to make the simulation more
realistic, the next subsection will tell us a different story.

\subsection{Comparison in Penna ageing model}

	Most organisms age, and thus we should compare sexual and asexual
reproduction in a model with ageing, where reproduction starts only after
a certain age. The Penna model of section 2 is the only one for which we
know of computer simulations for ageing and sex. The first comparisons of
MP with SX in this model were published by Bernardes (1997). More
recently, AS, AP, MP, HA and SX were simulated with it (Stauffer et al.
2000). We enlarge the range of possibilities by incorporating into it the
Jan parameter (Jan et al. 2000)  $\mu$ such that organisms with $\mu$ and
more mutations try to find a partner with whom they exchange genome (HA
and SX), while those with less than $\mu$ mutations use AP or MP.  One
counts only the mutations already set, up to the current age of each
individual, to make the choice.

	The simulated mixtures of reproduction were: MP-HA, MP-SX, AP-SX.
In the first case, the final population depends non-monotonically on $\mu$
showing a maximum for intermediate $\mu$, while in the two other mixtures
the behaviour is monotonic making the mixtures less interesting. Since $T$
mutations kill an individual, we have $0 \le \mu \le T$, with $\mu = 0$
describing pure HA and $\mu = T$ describing pure MP for the MP-HA mixture.
Fig.3 summarizes our main results.

	We see that SX (the small dots in the lower half) is by far the
worst, MP (+) and AS (line) give nearly identical results, AP (x) is
slightly worse than AS, and finally a mixture of HA and MP (stars) with
$\mu = 4$ gives the best results. Why then do males exist ?

	Some people claim men should eat less steaks and drink less
alcohol. Taken to the extreme, we assume the males eat nothing and are so
much smaller than the females that they consume no space. Some animals
followed this way of life long before us. Then their contribution to the
Verhulst dying probability $(N_m + N_f)/N_{max}$ mentioned in section 2
becomes negligible, and the dying probability is simplified to
$N_f/N_{max}$. With this change, the population for SX roughly doubles,
not surprisingly, and then SX is by far the best solution. The majority of
the present authors insist that this result is an artifact of the
assumptions and is no way to regulate their lives. In the evolution of sex
hundreds of million years ago it is difficult to imagine that together
with the mutation towards SX, immediately also the male body size became
much smaller and consumed much less food.

	Perhaps male aggressiveness plays a useful role in protecting the
children while reducing the male survival chances. Using the algorithm to
be described in section 4 for testosteron as an explanation for the lower
mortality of women compared to men, and female partner selection described
later in this section, Fig.4 shows that now SX is above MP or AS. This
child protection is already an effect outside the intrinsic genetic
effects discussed before. The following paragraphs discuss environmental
effects which can also justify SX over AS, in agreement with reality.

\bigskip

	Parasites have long been claimed to justify sexual reproduction,
since the greater genetic variety of the offspring gives the parasites
less chance to adapt to the host. An early computer simulation (Howard \&
Lively 1994) without ageing already showed sexual reproduction to die out
compared with asexual reproduction if no parasites are present (thus
similar to Redfield 1994), while in the presence of parasites sex can give
the better chance of survival (Howard \& Lively 1994). Within the more
realistic Penna model, including ageing, the parasite problem was studied
more recently by S\'a Martins (2000) with the same result: Parasites
justify sex.

	For this purpose, MP hosts, SX hosts and 1000 parasites were
simulated together in the same environment (S\'a Martins 2000)  
represented by $N_{max}$ for the sum of the populations. With probability
$1/10^4$, an individual can switch from MP to SX or back. The parasites
are represented by bit-strings. Each female host has contact with 60
parasites, and if one parasite agrees in its bit-string with that of the
female, this host loses her ability to procreate. The parasite bit-string,
on the other hand, is modified into the female bit-string if the parasite
meets the same bit-string for the second time.

	Starting with SX, the whole population changes to MP after a few
hundred time steps, if no parasites are present. In the presence of
parasites, however, starting with MP the whole population switches to SX
in an even shorter time (S\'a Martins 2000). Thus the well known greater
variety of SX (S\'a Martins \& Moss de Oliveira 1998; Dasgupta 1997)
compared to MP saves the sexual population from the attacks of parasites.

	Both papers (Howard \& Lively 1994; S\'a Martins 2000) were
motivated by observations of biologist Lively and his collaborators on
snails. They do not discuss if parasites already were plaguing the
presumably much smaller organisms nearly $10^9$ years ago when sex
appeared.

	Another reason to justify SX compared with MP are rapid changes in
the environment, to which natural evolution cannot be fast enough. SX and
HA lead across the species to a larger variety in the genome than AS or
MP, and after a catastrophe like the meteor killing the dinosaurs, the
species with a greater variety has a larger chance to contain a minority
of individuals adapted to the changed conditions. Fig.5, adapted from S\'a
Martins \& Moss de Oliveira (1998), compares MP with SX. First, MP gives
the higher population as in Fig.3, but when a sudden change in the
environment is introduced into the simulation, the SX population has a
higher chance to survive the catastrophe than MP.

	Another reason why genomic exchange between different individuals
may be better than AS, AP or MP is partner selection (Redfield 1994;
Stauffer et al. 1996).  Let us simulate in the sexual Penna model first a
birthrate of eight children per female and ``year''; then we reduce the
birth rate to four to account for males not getting pregnant; and in the
third step we assume that the females select as partners only healthy
males with few mutations. Some parameter region could be found, Fig.6,
where selection in step 3 gave a strong advantage over no selection (step
2), but this advantage was not enough to overcome the loss of half the
births compared with step 1. With the parameters of Fig.3 instead, the SX
data would shift upward to 18.5 million if only males with at most one
active mutation are selected.

	Thus, while the simpler models without ageing gave clear intrinsic
justifications for the existence of males, the more realistic ageing model
required parasites, catastrophes, or child protection for this purpose
(partner selection may help somewhat) and intrinsically slightly preferred
a HA-MP mixture over haploid asexual cloning. It remains to be seen what
SX simulations will give in other models (e.g. Onody 2000).

\section{Why women live longer and have menopause?}

	Men may be useless according to section 3.3, but why do they live
shorter than women, in the developed countries of the 20th century ?

	Many mutations limit life, and thus the higher mutation rate of
males compared with females (Redfield 1994) could be the reason of higher
male mortality. Simulations (Stauffer et al. 1996) showed that this is not
the case: The offspring is randomly either male or female and in both
cases inherits roughly the same mutations from the parents; for the same
reason, women are not killed by mutations after menopause (see below), in
contrast to Pacific Salmon (Moss de Oliveira et al. 1999; Penna et al.
1995).  Somatic mutations, which are not inherited and not given on to the
offspring, reduce the male mortality compared to the female one if the
rate of somatic mutations is higher for males than for females (Moss de
Oliveira et al. 1996).  Alternatively, females could be just more
resistant than males against diseases (Penna \& Wolf 1997). Mammalian
females have two X chromosomes, while the males have one X and one Y
chromosome such that all X mutations are dominant (Schneider et al. 1998).
Except at old age, these last three assumptions all lead to higher male
than female mortalities, as in reality (Fig.1), and where reviewed in our
book (Moss de Oliveira et al. 1999).

	A more recent idea was suggested to us by the medical researcher
Klotz and is related to male lifestyle (steaks and alcohol) caused by
testosteron (Klotz 1998; Klotz \& Hurrelmann 1998; Baulieu 1999).  This
hormone causes higher male aggressivity leading to death, as well as more
arteriosclerosis later in life. These bad effects were perhaps
counterbalanced earlier in human evolution by helping the males to defend
their families against predators or fellow men. This child protection by
males can in some other form also occur in other animals. Using this child
protection assumption together with sexual selection as described below,
the above SX results in Fig.4 were produced. And for today's humans, the
testosteron parameters of this child protection model could be chosen such
that the male mortality is about twice as high as the female mortality,
except at old age, in agreement with reality (Stauffer \& Klotz 2000;
Stauffer 2000).

	Presumably, the true reasons for the difference between the
mortalities of men and women is a combination of genetic and social
effects, as is shown by the variation from country to country within
Europe (Gjon\c ca et al. 1999). The XX-XY chromosome hypothesis (Schneider
et al. 1998)  is supported by the observation (Paevskii 1985)  that male
birds usually live longer than the females: For birds, the females have
two different and the males have two identical chromosomes, opposite to
mammals.

	(Technical remark: To simulate the effects of male testosteron
level in Fig.4, the Verhulst dying probability was increased, for males
only, by an amount proportional to the age-dependent testosteron level
$k(a)$ (Stauffer \& Klotz 2000; Stauffer 2000). On the other hand, the
probability of babies to survive was multiplied by a factor
min(const$\cdot k(a),2$) such that a too low testosteron level of the
father causes his babies to be killed by others. The function $k(a)$
evolved to an optimum shape through small heritable mutations in $k(a)$.)

\bigskip
	Menopause for women means an abrupt ceasing of their reproductive
function, while for men, andropause is a rather smooth decay with age.
Similar effects exist for the other mammals (though perhaps under
different names, which we ignore here). For Pacific Salmon, life ends for
males and females shortly after the end of reproduction of both (Penna et
al. 1995); why does the same effects not occur for women ?

	Pure genetic reasons (Stauffer et al. 1996) in the unmodified
Penna model, without child care, already allow women to survive menopause.
Conception decides randomly whether the new baby is a boy or a girl, and
the genome is the same apart from the difference in X and Y chromosomes.
Thus if all bits above reproductive age would be set equal to one for
women (as for Pacific Salmon), the men would also die at that same age of
menopause. To kill the women earlier than men, mother Nature would need a
longevity gene in the Y chromosome, which in reality contains little
genetic information. In this way, female survival after menopause is
consistent with the mutation-accumulation hypothesis in the Penna model.

	This consistency does not yet explain why menopause evolved. The
only simulation we are aware of explaining menopause (Moss de Oliveira et
al. 1999a)  introduces two new assumptions into the Penna model: a risk of
dying at birth for the mother which increases with the number of active
mutations and thus with age; and child care in the sense that young
children die if their mother dies. Then the maximum age for reproduction
was allowed to emerge from the simulation, instead of being put in fixed
at the beginning, by assuming it to be hereditary apart from small
mutations up or down. As a result, the distribution of the maximum age of
reproduction peaked at about 15 ``years'' whereas without child care its
maximum was at 32 years, at the oldest possible bit position.

\section{Other aspects}
	Geneticist S. Cebrat (priv. comm.) has criticized our crossover
method for the sexual Penna model as published in Moss de Oliveira et al.
(1999).  Since we split the bit strings at some randomly selected position
and then combine the first part of one bit-string with the second part of
the other bit-string, and since bit positions correspond to individual
age, we produce correlations for the mutations in consecutive ages. In
real DNA, the genes are not stored consecutively in the order in which
they become active during life. Thus it is better to select randomly one
subset of bits from one bit-string, and the other bits from the
complimentary bit-string. Simulations indicate no clear difference, Fig.7.

	Overfishing (Moss de Oliveira et al. 1995; Penna et al. 2000) and
the inheritance of longevity (de Oliveira et al. 1998) were simulated
using the asexual version of the Penna model.

	Although we have centered this work on the Penna model, it is not
the only way to go in ageing studies. It is an apropriate tool that allows
us to unravel the importance of mutation accumulation effects. The very
first models studied by physicists, summarized in Stauffer (1994), were
based on the antagonistic pleiotropy. They were inspired by the Partridge
and Barton review (Partridge \& Barton 1993). There they proposed a
constraint for the survival rates from babies to juveniles $J$ and
juveniles to adults $A$ as $ J + A^4 =1 $.  Because it has only two
parameters and three ages, the exponential increase of mortalities cannot
be observed. Some attempts of implementing antagonistic effects on
bit-string models have been done. Bernardes imposed an extra deleterious
mutation at advanced ages for a fraction of the population with higher
reproduction rates (Bernardes 1996).  Sousa and Moss de Oliveira, in a
more detailed study, have shown that the combined action of mutation
accumulation and antagonistic pleiotropy at defined ages can extend the
lifespan of a population (in preparation).  Sousa and Penna have
introduced a different strategy, where both sides of the antagonism are
present: good (bad) mutations at earlier (later) ages (in preparation).
The minimum age at reproduction is allowed to vary. The later the
individual reaches the sexual maturity, the more fertile it is. There is a
clear compromise to postpone the maturity (and consequently to decrease
the integrated fertility) against to be more exposed to death by
competition or action of bad mutations. Preliminary results suggest that
except for unrealistic handicaps on the fertility for later maturity,
natural selection drives the populaton to earliest maturity (see also
Medeiros et al. 2000).

\section{Summary}
	The computer simulations of biological ageing, mostly using the
Penna model, could explain nicely the roughly exponential increase of
mortality functions with age, the existence of menopause, and (with less
clarity) the existence of other forms of reproduction besides asexual
cloning of haploid genomes. We speculate that from this asexual way,
mother Nature may have evolved via apomictic and/or meiotic
parthenogenesis towards hermaphroditism, and only later separated the
population into males and females because of external or social reasons,
as simulated. Finally, menopause appeared because of the need for child
care by the mother and the risks for her associated with giving birth
later in life. For practical applications, simulations suggested not to
catch young fish, or young and old lobsters, in order to maximize the
catch (Moss de Oliveira et al. 1995; Penna et al. 2000).

\bigskip
\parindent 0pt
\bigskip

\centerline{\bf REFERENCES}
\bigskip

\bigskip

{\sc Azbel MYa}. 1994. Universal biological scaling 
and mortality. {\it Proc. Natl. Acad. Sci. USA} {\bf 91}: 12453-12457.
\bigskip

{\sc Bak P}. 1997. {\it How Nature Works: the Science of Self-Organized
Criticality}, Oxford University Press.
\bigskip

{\sc Baulieu EE}. 1999. Le vieillissement est-il soluble dans 
les hormones? {\it La Recherche} {\bf 322}: 72-74. 
\bigskip

{\sc Bernardes AT}. 1995. Mutational meltdown in large sexual 
populations. {\it J. Physique} I {\bf 5}: 1501-1515.
\bigskip

{\sc Bernardes AT}. 1996. Strategies for reproduction and 
ageing. {\it Ann. Physik} {\bf 5}: 539-550.
\bigskip

{\sc Bernardes AT}. 1997. Can males contribute to 
the genetic improvement of the species?  
{\it J. Stat. Phys.} {\bf 86}: 431-439.
\bigskip

{\sc Darwin C}. 1859. {\it On the Origin of Species by Means of
Natural Selection}, Murray, London.
\bigskip

{\sc Dasgupta S}. 1997. Genetic crossover vs. cloning by computer simulation.
{\it Int. J. Mod. Phys.} C {\bf 8}: 605-608.
\bigskip

{\sc de Oliveira PMC}. 1991. {\it Computing Boolean Statistical
Models}, World Scientific, Singapore/London/New York.
\bigskip

{\sc de Oliveira PMC}. 2000. Why do evolutionary 
systems stick to the edge of chaos. {\it Theory in Biosci.}: in press.
\bigskip

{\sc de Oliveira PMC, Moss de Oliveira SM, Bernardes AT \& Stauffer 
D}. 1998. {\it Lancet} {\bf 352}: 911-912. 
\bigskip

{\sc Excoffier L}. 1997. Ce que nous dit la
genealogie des genes. {\it La Recherche} {\bf 302}: 82-84.
\bigskip

{\sc Gjon\c ca A, Tomassini C \& Vaupel JW}. 1999. 
Pourqoi les femmes survivent aux hommes? 
{\it La Recherche} {\bf 322}: 96-99.
\bigskip

{\sc Holbrook NJ, Martin GR \& Lockshin RA}. 1996. {\it Cellular 
Ageing and Death}, Wiley-Liss, New York.
\bigskip

{\sc Howard RS \& Lively CM}. 1994. Parasitism, 
mutation accumulation and the maintenance of sex. 
{\it Nature} {\bf 367}: 554-557 and {\bf 368}: 358 (Erratum).
\bigskip

{\sc Jacquard A}. 1978. {\it \'Eloge de la Diff\'erence: la
G\'en\'etique et les Hommes}, \'Editions du Seuil, Paris.
\bigskip

{\sc Jan N, Moseley L \& Stauffer D}. 2000. A hypothesis 
for the evolution of sex. {\it Theory in Biosci.} {\bf 119}: 166-168.
\bigskip

{\sc Kauffman SA}. 1993. {\it Origins of Order: Self-Organization and
Selection in Evolution}, Oxford University Press, New York.
\bigskip

{\sc Kauffman SA}. 1995. {\it At home in the Universe}, Oxford 
University Press, New York.
\bigskip

{\sc Klotz T}. 1998. {\it Der fr\"uhe Tod des starken Geschlechts}, 
Cuvillier, G\"ottingen.
\bigskip

{\sc Klotz T \& Hurrelmann K}. 1998. Adapting the health care system 
to the needs of the aging male. {\it The Aging Male} {\bf 1}: 20-27.
\bigskip

{\sc Lynch M \& Gabriel W}. 1990. Mutation load and the survival of 
small populations. {\it Evolution} {\bf 44}: 1725-1737.
\bigskip

{\sc Medeiros G, Idiart MA \& de Almeida RMC}. 2000. 
Selection experiments in the Penna model for biological aging.  
{\it Int. J. Mod. Phys.} C {\bf 11}: No. 7. 
\bigskip

{\sc Moss de Oliveira, Penna TJP \& Stauffer D}. 1995. 
Simulating the vanishing of northern cod fish.
{\it Physica} A {\bf 215}, 298-304.
\bigskip

{\sc Moss de Oliveira S, de Oliveira PMC \& Stauffer D}. 1996. Ageing 
with sexual and asexual reproduction: Monte Carlo simulations of
mutation accumulation. {\it Braz. J. Phys.} {\bf 26}: 626-630.
\bigskip

{\sc Moss de Oliveira S, de Oliveira PMC \& Stauffer D}. 1999.
{\it Evolution, Money, War and Computers}, Teubner, Leipzig.
\bigskip

{\sc Moss de Oliveira S, Bernardes AT \& S\'a Martins JS}. 1999a. 
Self-organisation of female menopause in populations with child care 
and reproductive risk. {\it Eur. Phys. J.} B {\bf 7}: 501-504.
\bigskip

{\sc \"Or\c cal B, T\"uzel E, Sevim V, Jan N \& Erzan A}. 2000. 
Testing a hypothesis for the evolution of sex. 
{\it Int. J. Mod. Phys.} C {\bf 11}: 973-986; and also DeCoste C \&
Jan N, priv. comm.
\bigskip

{\sc Onody RN}. 2000. The Heumann-H\"otzel Model revisited. Talk O-24 
at FACS 2000, Macei\'o, Brazil.
\bigskip

{\sc Paevskii VA}. 1985. {\it Demography of Birds} (in Russian), Nauka, 
Moscow. 
\bigskip

{\sc Pamilo P, Nei M \& Li WH}. 1987. Accumulation of mutations in 
sexual and asexual populations. {\it Genet. Res., Camb.} {\bf 49}: 135-146.
\bigskip

{\sc Partridge L \& Barton NH. 1993}. Optimality, 
mutation and the evolution of ageing. {\it Nature} {\bf 362}: 305-311.
\bigskip
 
{\sc Penna TJP}. 1995. A bit-string model for biological 
ageing. {\it J. Stat. Phys.} {\bf 78}: 1629-1633.
\bigskip

{\sc Penna TJP \& Moss de Oliveira S}. 1995. Exact results of the
bit-string model for catastrophic senescence. {\it J. Physique} I {\bf 5}: 1697-1703.
\bigskip

{\sc Penna TJP, Moss de Oliveira S \& Stauffer D}. 1995. Mutation 
accumulation and the catastrophic senescence of the Pacific salmon. 
{\it Phys. Rev.} E {\bf 52}: R3309-R3312.  
\bigskip

{\sc Penna TJP \& Stauffer D}. 1996. Bit-string ageing model and
German population. {\it Zeits. Phys.} B {\bf 101}: 469-470.
\bigskip

{\sc Penna TJP \& Wolf D}. 1997. Computer simulation 
of the difference between male and female death rates. 
{\it Theory in Biosc.} {\bf 116}: 118-124.
\bigskip

{\sc Penna TJP, Racco A \& Sousa AO}. 2000. Can microscopic models for 
age-structured populations contribute to Ecology? 
Talk IT-11 at FACS 2000, Macei\'o, Brazil (to appear in {\it Physica A}).
\bigskip

{\sc Redfield RJ}. 1994. Male mutations and the 
cost of sex for males. {\it Nature} {\bf 369}: 145-147.
\bigskip

{\sc Rose MR}. 1991. {\it Evolutionary Biology of
Aging}, Oxford University Press, New York.
\bigskip

{\sc S\'a Martins JS}. 2000. Simulated coevolution 
in a mutating ecology. {\it Phys. Rev.} E {\bf 61}: 2212-2215.
\bigskip

{\sc S\'a Martins JS \& Moss de Oliveira S}. 1998. 
Why sex - Monte Carlo simulations of survival after catastrophes. 
{\it Int. J. Mod. Phys.} C {\bf 9}: 421-432.
\bigskip

{\sc Schneider J, Cebrat S \& Stauffer D}. 1998. 
Why do women live longer than men? A Monte Carlo Simulation 
of Penna-type models with X and Y cromossomes. {\it Int. J. Mod. Phys.} C {\bf 9}: 
721-725.
\bigskip

{\sc Schr\"odinger E}. 1944. {\it What is Life?}, Cambridge
University Press, Cambridge.
\bigskip

{\sc Sousa AO \& Moss de Oliveira S}. 1999. 
High reproduction rate versus sexual fidelity. {\it Eur. Phys. J.} B {\bf 10}: 781-785.
\bigskip

{\sc Stauffer D}. 1994. Monte Carlo 
simulations of biological ageing. {\it Braz. J. Phys.} {\bf 24}: 900-906.
\bigskip

{\sc Stauffer D}. 1999. Why care about sex? Some Monte Carlo 
justification. {\it Physica} A {\bf 273}: 132-139.
\bigskip

{\sc Stauffer D}. 2000. Self-organisation of testosterone level 
in the Penna-Klotz ageing model. {\it Theory in Biosciences}, in press.
\bigskip

{\sc Stauffer D \& Aharony A}. 1994. {\it Introduction to 
Percolation Theory}, Taylor and Francis, London.
\bigskip

{\sc Stauffer D, de Oliveira PMC, Moss de Oliveira S \&  
Zorzenon dos Santos RM}. 1996. Monte Carlo simulations of sexual
reproduction. {\it Physica} A {\bf 231}: 504-514.
\bigskip

{\sc Stauffer D \& Klotz T}. 2000. The mathematical point of view: The 
sex-specific life expectancy and the influence of testosterone in an 
aging simulation model and its consequences for prevention.
Submitted to {\it The Aging Male}.
\bigskip

{\sc Stauffer D, S\'a Martins JS \& Moss de Oliveira 
S}. 2000. On the uselessness of men - Comparison of sexual and 
asexual reproduction. {\it Int. J. Mod. Phys.} C {\bf 11}: No. 7.
\bigskip

{\sc Vaupel JW, Carey JR, Christensen k, Johnson TE, 
Yashin AI, Holm NV, Iachine IA, Kanisto V, Khazaeli AA, Liedo P, Longo
VD, Zeng Y, Manton KG \& Curtsinger JW}. 1998. Biodemography of
longevity. {\it Science} {\bf 280}: 855-860.
\bigskip

{\sc Wachter KW \& Finch CE}. 1997. {\it Between Zeus and the 
Salmon. The Biodemography of Longevity}, National Academy Press, Washington DC.
\bigskip

{\sc Wilson KG}. 1971. Renormalization group and 
critical phenomena I. Renormalization group and the Kadanoff scaling 
picture. {\it Phys. Rev.} {\bf B4}: 3174-3183.
\bigskip

{\sc Wilson KG \& Kogut J}. 1974. The renormalization group and the
$\epsilon$ expansion. {\it Phys. Rep.} {\bf 12C}: 75-200.
\bigskip

{\sc Wilson KG}. 1979. Problems in physics with many scales of 
length. {\it Sci. Am.} {\bf 241}: 140-157.

\newpage
\section {Figure Captions}

\parindent 0pt

Fig.1: Male (+) and female (x) mortality functions in the USA, 1991-1995; from 
J.R. 
Wilmoth's Berkeley Mortality Database demog.berkeley.edu/wilmoth/. The straight
line through the male data indicates the exponential increase with age (Gompertz
law). These mortality functions are defined as $-d\ln S(a)/da$ where $S(a)$ is
the probability to survive up to an age of $a$ years. 

\medskip
Fig.2: Schematic representation of the genomic changes for AS, AP, MP (from 
left to right) in part a
and for SX in part b. (The diagram for SX is also valid for HA, except that
for HA all individuals can reproduce.)

\medskip
Fig.3: Comparison of populations, versus number of iterations or ``years'', for
(from below) SX, AP, MP. The highest data refer to a mixture of HA and MP
with $\mu = 4$; nearly the same results are obtained for $\mu = 3$ and 5.
The data for AS (line) overlap with those of MP (+), while AP(x)
fluctuates around slightly lower values. For SX we show the sum
of males and females. Threshold $T=9$, 4 births per year and 
per female above minimum reproduction rate of 8, one mutation per 
string of 32 bits at birth, $N_{max}$ = 80 million is about four times larger 
than the actual populations. 

\medskip
Fig.4: Comparison of populations, versus number of iterations, for SX with 
child protection (+,x), and AS (line). (For the data marked by x, females select only male 
partners with at most one active bad mutation.) $N_{max}$ = 5 million; otherwise parameters 
as in Fig.3.

\medskip
Fig.5: Comparison of MP (lines) with SX (crosses)
population before and after a sudden change in the environment. 
For SX, the female birth rate of two was the same as for all MP individuals. 
($N_{max}= 400000, \; T=3, \; d = 5, \; R = 10$, one mutation per bit-string.)

\medskip
Fig.6: Can partner selection overcome the loss of half the births ? The top
data show step 1, the bottom data step 2, and the middle data step 3
(see text): Selection helps, but not enough.

\medskip
Fig.7: Comparison of traditional ordered crossover (x, squares) with better
random crossover (+, stars) showing little difference. The Verhulst deaths
appear at all ages (+, x) or only at birth (stars, squares). 
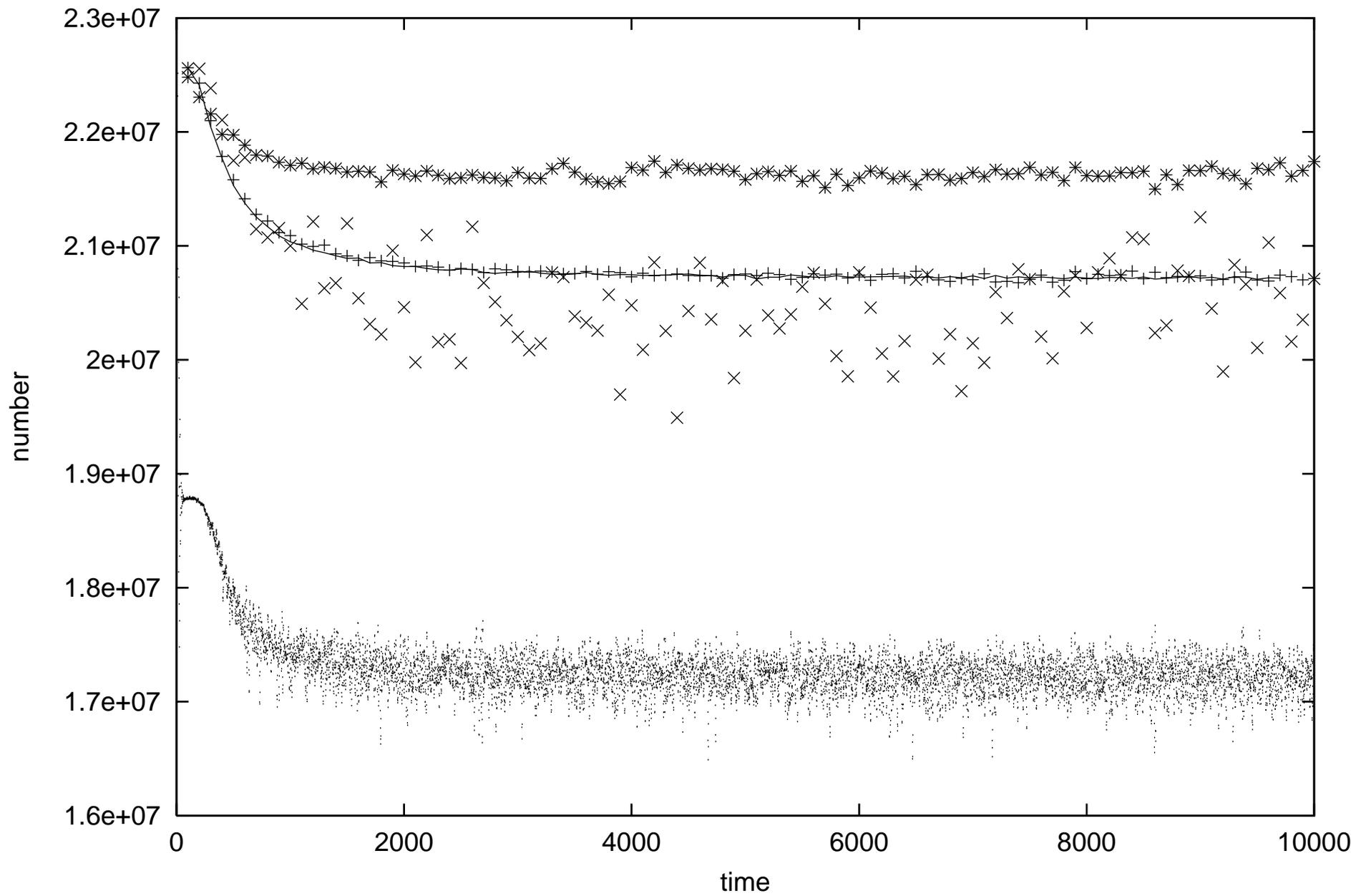
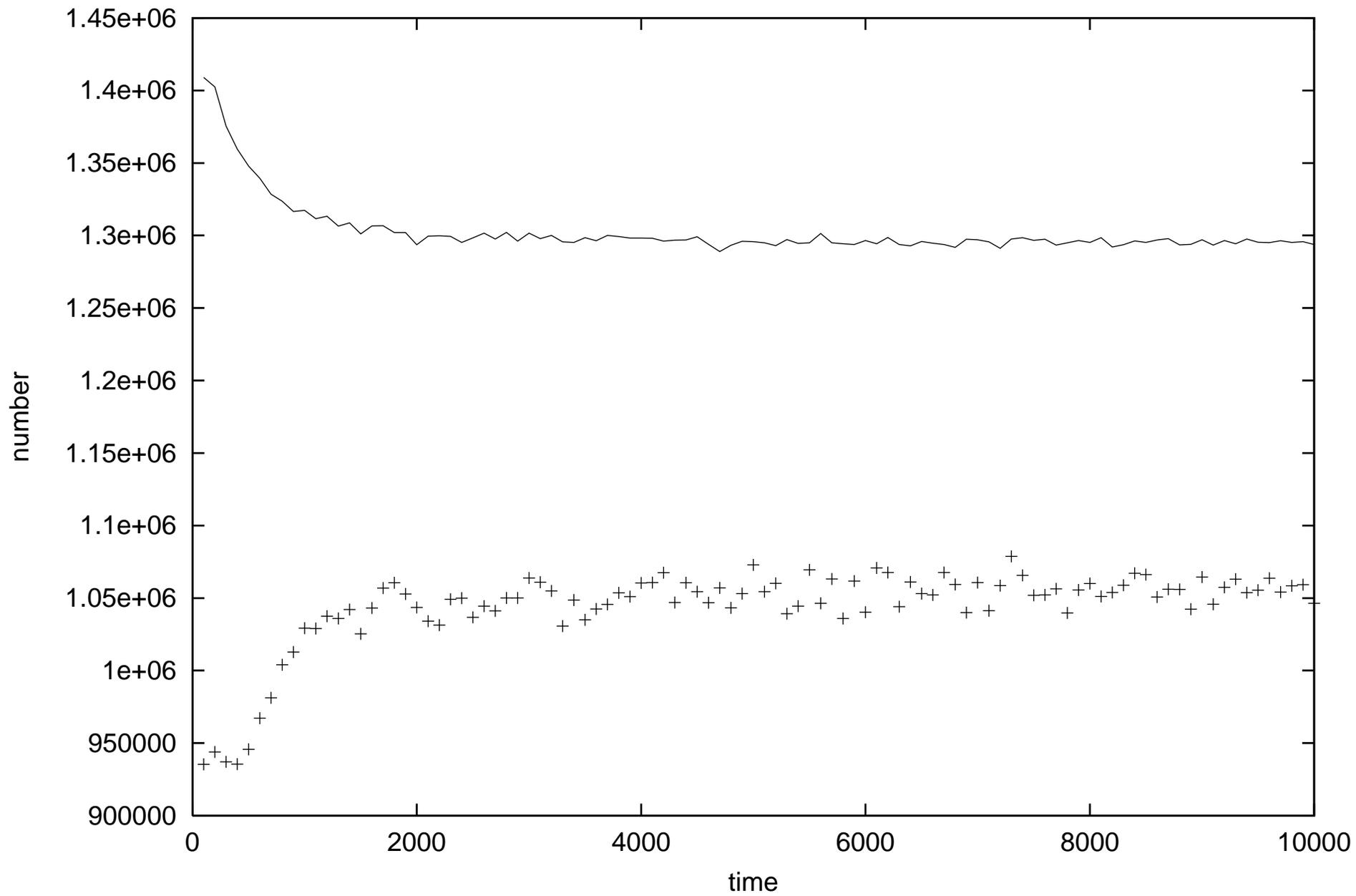
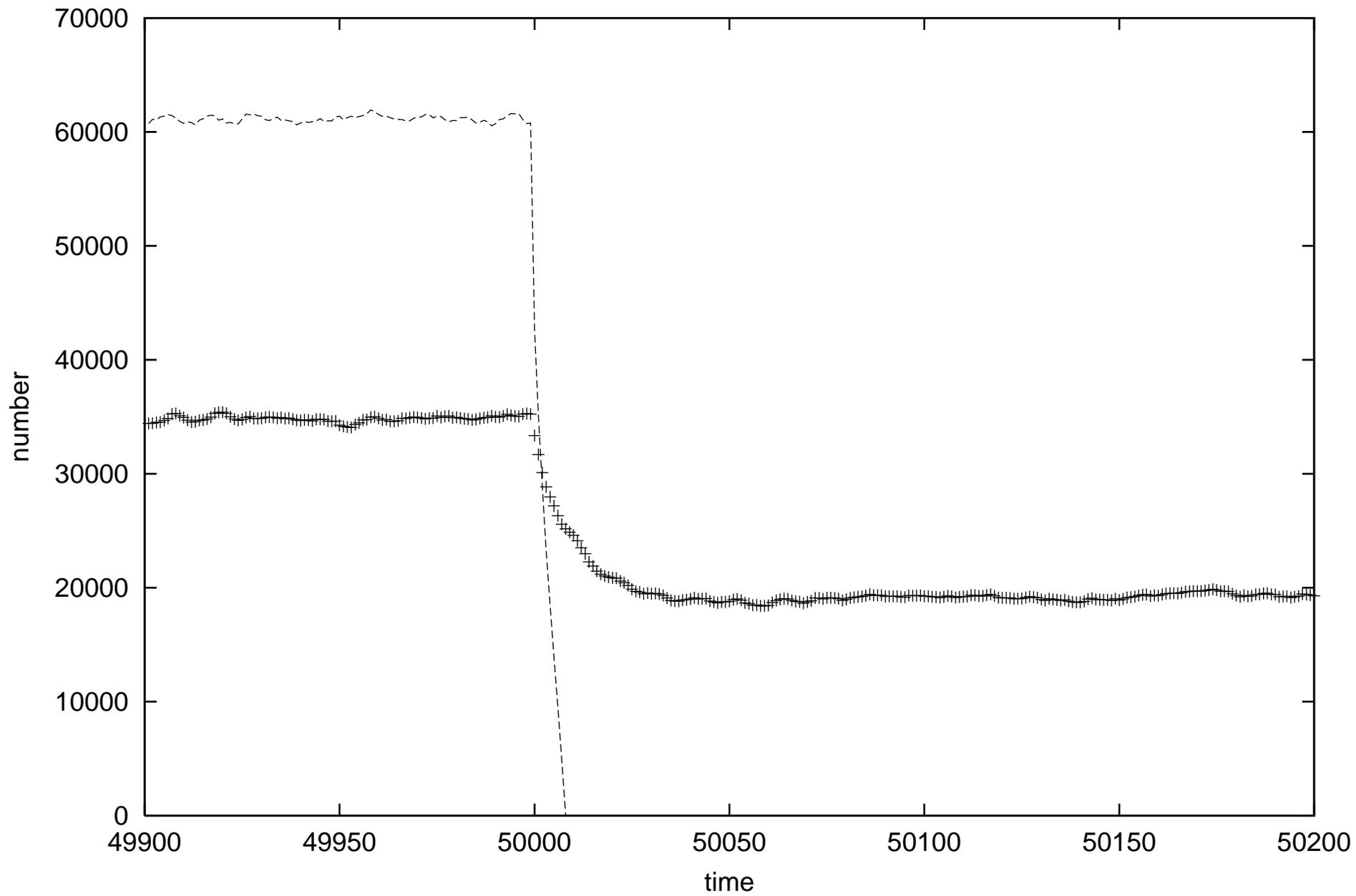
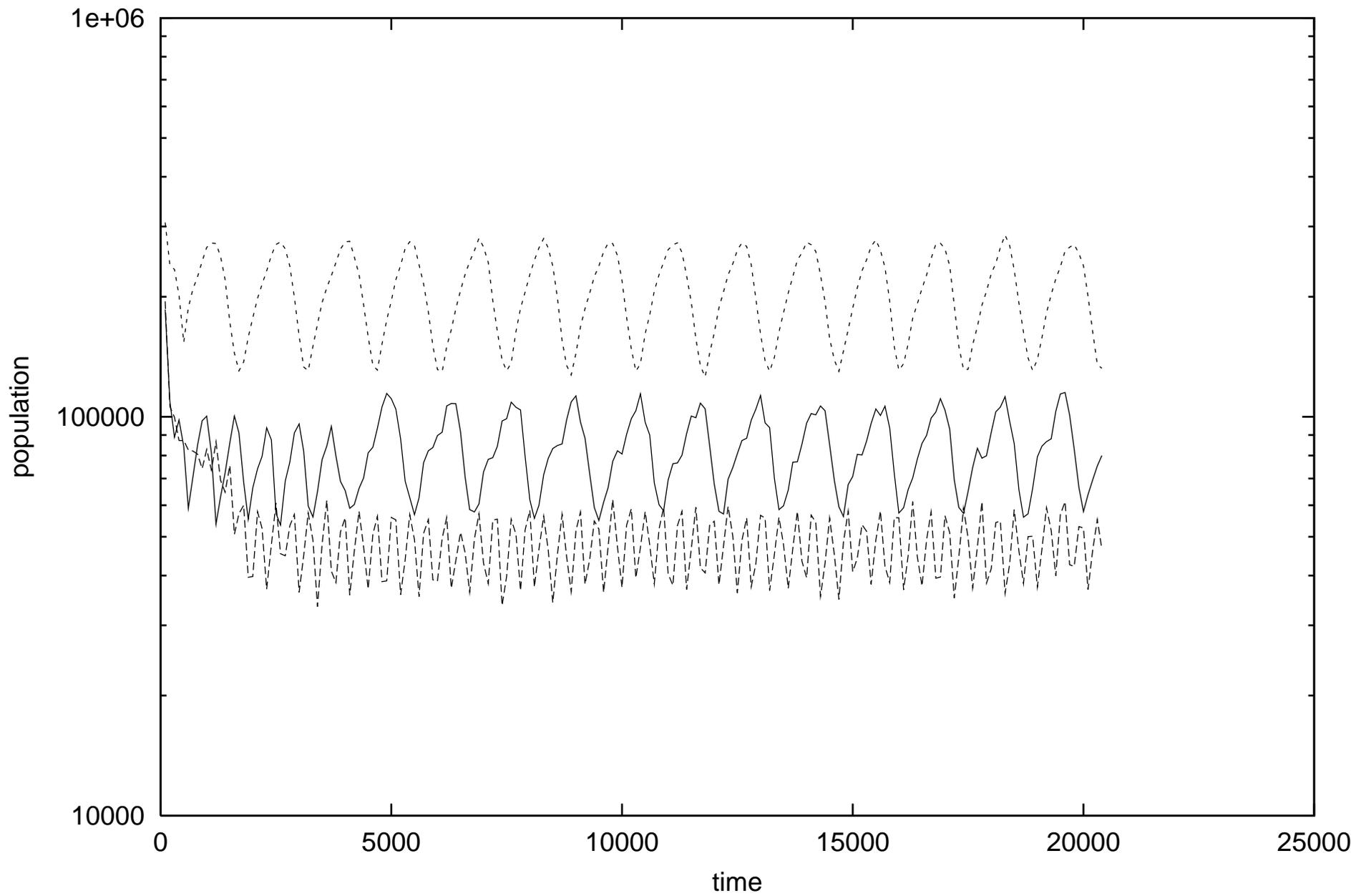
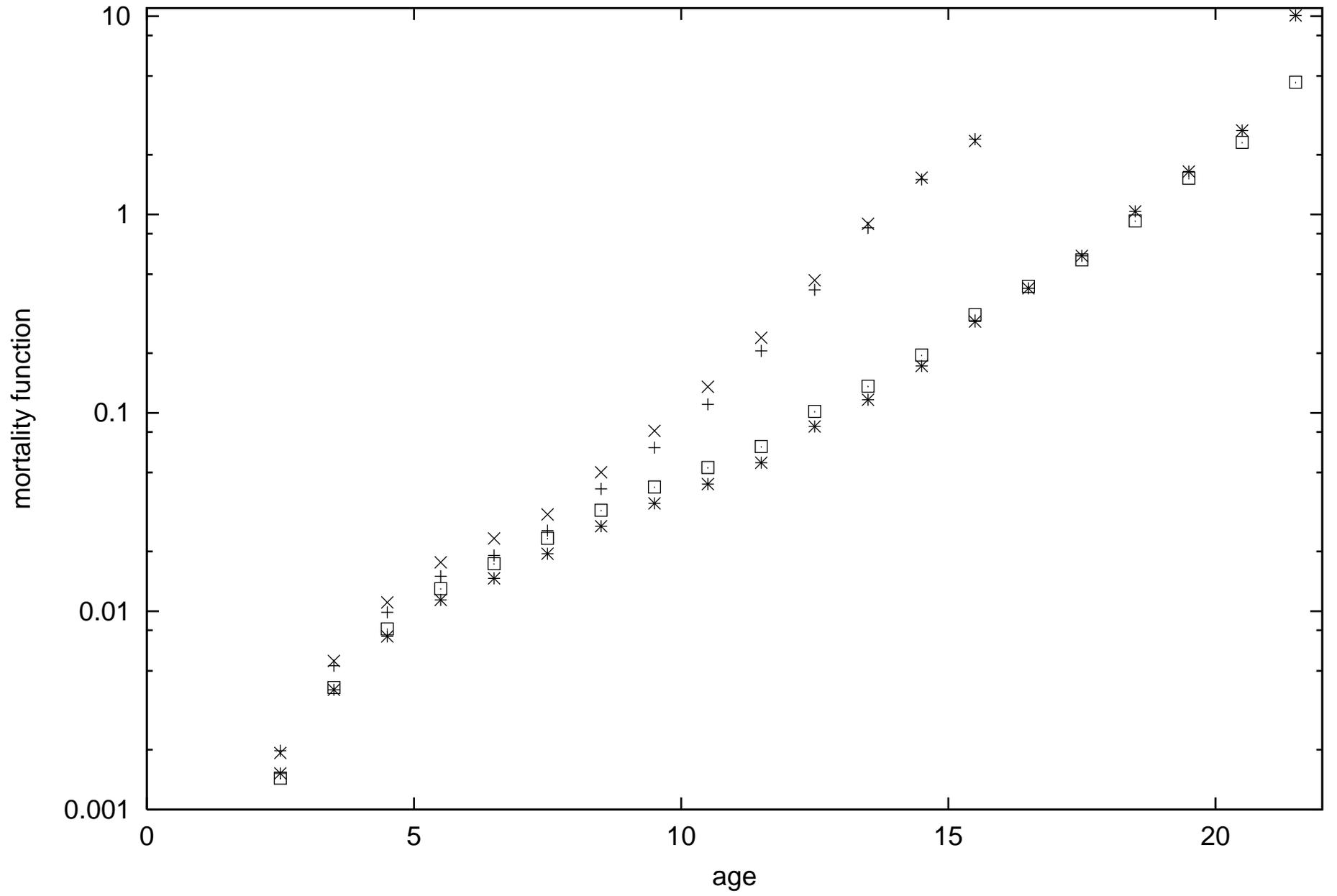
\end{document}